\documentstyle[aps,pre,epsf]{revtex}
\begin{document}
\draft
\twocolumn[\hsize\textwidth\columnwidth\hsize\csname
@twocolumnfalse\endcsname 
\title{\bf A Simple Model for Plastic Dynamics of a Disordered
Flux Line Lattice}
\author{ 
Kevin E. Bassler$^1$, Maya Paczuski$^{2,1}$, and Ernesto Altshuler$^{3,4}$}
\address{
$~^1$ Department of Physics, University of Houston, Houston TX 77204-5506 \\
$~^2$ Department of Mathematics, Imperial College of Science, 
Technology, and Medicine, London, UK SW7 2BZ\\
$~^3$ Texas Center for Superconductivity, University of Houston, Houston TX 
77204-5506 \\
$~^4$ Superconductivity Laboratory, IMRE-Physics Faculty, University of Havana, 
10400 Havana, Cuba
}
\date{\today}

\maketitle 

\begin{abstract}

We use a coarse-grained model of superconducting vortices driven
through a random pinning potential to study the nonlinear
current-voltage (IV) characteristics of flux flow in type II
superconductors with pinning.  In experiments, the IV relation
measures flux flow down a flux density gradient.  The work presented
here treats this key feature explicitly.  As the vortex repulsion
weakens, the vortex pile maintains a globally steeper slope,
corresponding to a larger critical current, for the same pinning
potential.  In addition, the magnitude of the peak in the differential
resistance falls as the resistance peak shifts to higher currents.
The model also exhibits so-called ``IV fingerprints'', and crossover
to Ohmic (linear) behavior at high currents.  Thus, many of the
varieties of plastic behavior observed experimentally for soft flux
line systems in the ``peak regime'' are reproduced in numerical
simulations of the zero temperature model.  This model describes a
two-dimensional slice of the flux line system at the scale of the
London length $(\lambda)$.  It does not include any degrees of freedom
at scales much smaller than $\lambda$, which are required to specify
the degree of disorder in a flux line lattice.  Instead, the nonlinear
transport behaviors are related to the self-organized, large scale
morphologies of the vortex river flow down the slope of the vortex
pile.  These morphologies include isolated filamentary channels, which
can merge at higher flow rates to make a braided river, and eventually
give uniform flow at even higher flow rates.  The filamentary
structure is associated with an IV characteristic that has concave, or
positive, curvature. The braided river is associated with the peak in
the differential resistance, where the curvature of the IV relation
changes to convex.  The transition to Ohmic behavior comes about as
the braided river floods when it cannot absorb a higher level of
flow. We propose that these self-organized morphologies of flux flow
down a flux gradient govern the various plastic flow behaviors,
including nonlinear IV characteristics, observed in type II
superconductors with random pinning.
\end{abstract}

{PACS numbers: 74.60Ge, 74.60Jg, 64.60Ht, 62.20.Fe}
\vskip2pc]

\narrowtext
\section{Introduction}

Collective transport in disordered media is a widespread and poorly
understood phenomena.  A great deal of experimental and theoretical
effort in this area has been devoted to studying the nonlinear
dynamics of the disordered flux line lattice (FLL) in type II
superconductors.  The FLL exhibits a threshold behavior due to the
competition between pinning and flux line repulsion
\cite{vort-rev,vort-rev2}.  In response to a force, such as that
associated with a transport current, the three dimensional FLL can
move smoothly via elastic deformations, maintaining its integrity and
order.  However, in another regime the FLL deforms plasticly.  In that
regime disorder becomes more important, and the moving FLL manifold
tears as some flux lines (in two-dimensions, vortices) move while
others do not.  As a result, the flow pattern breaks up in a
nonuniform way.  It is generally believed that the microscopic
structure of the FLL, or defects in it, is fundamental to transport
behavior both in the plastic and elastic regimes.

Part of the attention to the dynamics of a moving FLL has been
motivated by interests in possible, exotic phase transitions and
glassy phases, melting and other complicated scenarios associated with
structural order in the FLL.  Most experiments, however, are
essentially transport studies, and are, quoting Higgins and
Bhattacharya\cite{HB96}, ``notoriously ill-suited for the study of
thermodynamic phase transitions.  These experiments yield direct
information only about the mobility of flux lines, i.e. on dynamics
and pinning, which would then have to be connected, through highly
model-dependent ways, to the structure of the state they pertain to.''

 Here, we show that many of the empirical results found in transport
 studies of plastic flux flow in type II superconductors may be
 obtained with an extremely simple model\cite{kevin}.  It has been
 recognized for many years that, in the presence of pinning, magnetic
 flux in type II superconductors forms a pile with an overall, global
 slope, akin to a sandpile.  Penetration of magnetic flux into
 superconductors driven solely by the flux density gradient has been
 described using molecular dynamics (MD) simulations 
\cite{gdMD0,gdMD1,gdMD2,gdMD3,gdMD4,gdMD5}, and
 by the model used here \cite{kevin}. However, the flux gradient has
 not been taken into account in any previous numerical simulation
 studies of the current-voltage characteristic.  One possible reason
 is that previous numerical studies of flux motion at the scale of the
 vortex cores have not been able to reach a sufficiently large system
 size.  The Bassler-Paczuski (BP) model \cite{kevin}, on the other
 hand, is a coarse-grained model and describes the magnetic flux
 dynamics  at the much larger scale of the London
 length, making the large system size limit much more accessible.

 The numerical simulations presented here of the BP model \cite{kevin}
show that nonlinear behaviors, characteristic of experimental
transport measurements of plastic flow in superconductors, arise as a
result of vortex flow down a vortex density gradient.  The effect of
the transport current is modelled by a shift in boundary conditions,
which leads to a generalized ``tilt'' of the vortex pile.  Eventually
the ``tilt'' is sufficiently great so that some vortices can flow down
the pile in the steady state, leading to the onset of a finite
voltage.  The vortex flow forms a variety of river morphologies
depending on the interaction strength between the vortices, compared
to pinning, and the overall rate of flow.  The flow patterns of
magnetic flux are self-organized together with the magnetic flux
profile, which is the substrate on which the flow takes place.  Since
the BP model does not contain any detailed information on the
positions of the vortex cores at the micro-scale of the FLL, it cannot
exhibit any structural ordering or disordering behavior.  

Although the BP model can also be studied at finite temperatures, here
we use the zero temperature limit where thermal fluctuations of the
flux motion may be ignored.  The zero temperature approximation seems
reasonable to describe the plastic transport dynamics of the low
temperature superconductors and, perhaps, some aspects of the high
temperature superconductors as well.  Also, we consider the limit
where the depairing current density, $j_0$, is extremely large
compared to the critical current density, $j_c$.  This corresponds to
the so called ``weak pinning'' regime.

 We argue that many of the varieties of collective transport dynamics
 observed in superconductors may be generic to repulsive particle
 systems driven through a disordered media.  These behaviors are
 directly related in our coarse-grained model to large scale
 morphologies of flow down a vortex density gradient and changes from
 filamentary strings, to a braided river, to uniform flow at high
 applied currents.  Since the BP model arguably contains the essential
 physics of the disordered flux line system, at a coarse-grained level, and
 reproduces a wide variety of experimental results on transport
 properties, we propose that these self-organized, large scale flow
 morphologies are also governing the nonlinear, plastic dynamics in the actual
 physical system:  flux lines driven through a superconductor with
a disordered pinning landscape by an applied transport current.

\subsection{Summary}

In the next section,  we focus
attention  on an extensive study of $2H-NbSe_2$ as
summarized by Higgins and Bhattacharya\cite{HB96}.  The reader
may already compare Fig.~1 and Fig.~2, which are schematic reproductions
of  experimental results, with Figs.~6 and
7, which are results from numerical simulations presented in this work.
 Section III presents a short review
the standard theoretical picture, associated with the work of
Shi and Berlinsky\cite{shi}, on tearing of the FLL and the resultant
nonlinear IV as found via molecular dynamics (MD) 
simulations\cite{jensen1,jensen2,brass,gronbechjensen,nori1,nori2,kolton}.  
These MD simulations do not, however, take into account a vortex density
gradient,  and do not agree with some
 the varieties of plastic dynamics observed
in experiments.  Section IV describes the original coarse-grained
model, first used to describe vortex avalanches\cite{kevin} and vortex
rivers\cite{braidedrivers} as seen in experiments 
\cite{field,lorentz}, and explains the
exact model used here to describe the transport experiments
summarized in Section II.  To eliminate all potentially spurious sources of 
noise,
 a completely deterministic variant of the original
BP model is used.  To describe the IV experiments, we apply a
shift in boundary conditions or generalized ``tilt'' of the vortex
pile to represent the effect of a transport current.  The resultant
vortex flow represents the measured voltage. Section V contains
the main numerical results and comparisons with both
experiments and MD simulations.  The last section summarizes our
conclusions.

\section{Brief Summary of transport measurements}

In a series of papers, Bhattacharya and
Higgins\cite{HB96,bh-prl93,bhfingerprints,marley} describe experiments
on the nonlinear transport properties of the FLL in the anisotropic
superconductor $2H-NbSe_2$.  In this system, the pinning is extremely
weak ($j_c/j_0 \sim 10^{-6}$), the lattice is well-formed and a robust
``peak effect'' occurs slightly below $H_{c2}$.  Since the London
length, $\lambda$, is much less than the film thickness, the system
operates in the three dimensional regime, where the flux line
interactions decay exponentially for lengths larger than $\lambda$.
They use the strong magnetic field dependence of the critical current
(which they interpret in terms of a changing rigidity of the FLL) to
explore the crossover between different type of dynamics including
``elastic'' flow, ``plastic'' flow, and ``fluid'' flow.  In this
regard, the material they study is an ideal experimental system,
allowing the exploration of very different regimes in a
well-controlled manner.

Near the upper critical magnetic field,  they observe a
 pronounced peak in the pinning force.  This is referred to as a
 ``peak effect''.  (It should be distinguished from the peak in the
differential resistance to be described later.)
Equivalently, the critical current, $I_c$, where
 some flux lines start to move, increases as the external
 magnetic field, $H$, increases.  For $I<I_c$ the entire flux system is
 pinned, while for $I>I_c$, some magnetic flux lines flow across the
 sample leading to the onset of a finite voltage, $V$.  We will focus
 mostly on results associated with the current-voltage (IV) relation
 as the external magnetic field, $H$, is varied.

\begin{figure}
\narrowtext
\epsfxsize=3.0truein
\centerline{ \epsffile{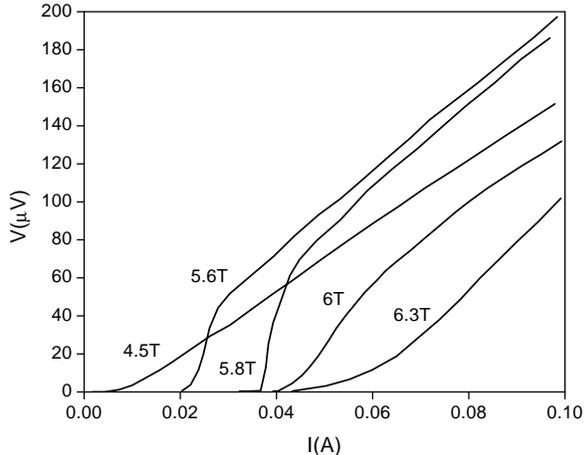} }
\caption{ 
Schematic figure illustrating the experimental IV measurements of
Bhattacharya and Higgins in $2H-NbSe_2$
at 4.2 K with varying applied field. After Fig.~1b of 
Ref.~\protect\onlinecite{bh-prl93}.
}
\end{figure}

As shown in Fig.~1b of Ref.~\onlinecite{bh-prl93}, in the ``peak regime''
the IV characteristics of the
superconductor vary enormously.  For clarity, these
experimental observations are illustrated
schematically in Fig.~1.  First, on
increasing $H$, the critical current, $I_c$,
increases.  This is the ``peak effect''.
Second, below some threshold magnetic field, the IV curves
always rise concave upward from $I_c$.  This is the generic form of
the IV for an FLL that is usually reported in the literature, and it
is associated with an ``elastic'' regime.  Third, when the external
magnetic field enters the ``peak regime'', the IV curves change
drastically, starting as concave upwards but then bending over after a
pronounced inflection point associated with a change of curvature.  In
the plastic regime, one obtains a characteristic S-shape IV curve.
Close to onset, $I_c$, it is concave upward, but then bends over as
$I$ increases further, saturating to a finite slope at large currents.
Between the S-shaped and elastic IV curves may be a special curve
which is always convex for $I>I_c$.  This appears at approximately
5.8T in the experiments where the inflection point has moved to onset.
Finally, above another magnetic field value, the inflection point is
at currents larger than those used in the experiments, and there is no
saturation in the slope of the IV curves that can be observed.  The
numerical simulations presented here
reproduce the entire progression of these curves in the
plastic regime and their changes as a parameter in the model,
representing vortex interactions, is varied.  It does not reproduce
the behavior in the ``elastic'' regime, as explained later.  The
elastic regime is also not observed in two-dimensional MD simulations
of driven vortices near the onset of flow.

\begin{figure}
\narrowtext
\epsfxsize=3.0truein
\centerline{ \epsffile{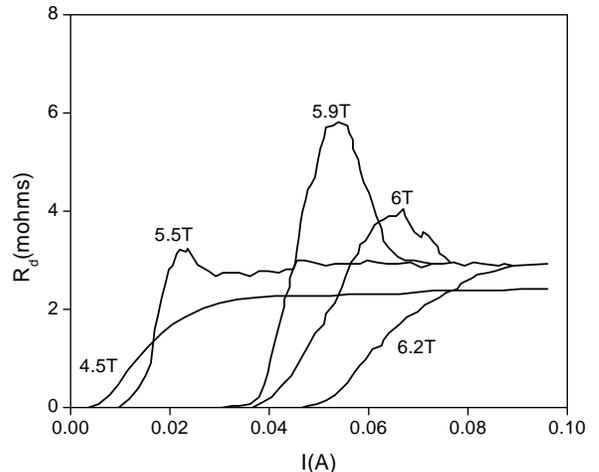} }
\caption{ 
Schematic figures illustrating the experimental Differential resistance 
measurements corresponding to the IV measurements shown in Fig.~1.
After Fig.~1c of 
Ref.~\protect\onlinecite{bh-prl93}.}
\end{figure}

Experimental measurements of the differential resistance, $R=dV/dI$,
reveal a peak in $R$, corresponding to the inflection point for the
S-shaped IV curves.  {\it The magnitude of this peak diminishes 
as the external magnetic field approaches $H_{c2}$, and the position of
the peak shifts to higher currents.}  
Except for the largest magnetic fields used, the
differential resistance eventually saturates at constant values
at large enough currents, indicating Ohmic or fluid like behavior for
sufficiently high driving.  Also, in the plastic regime corresponding
to the S-shaped IV curves, the usual scaling ansatz associated with
dynamic critical phenomena $V \sim (I-I_c)^{\beta}$ does not appear to
hold. These
experimental observations are illustrated
schematically in Fig.~2. The numerical simulations presented here
 reproduce this precise pattern of behavior for the
differential resistance in the ``peak'' regime, including the changing
position (to higher currents) and decreasing magnitude of the peak
resistance as vortex interactions weaken, and the saturation to Ohmic
or fluid-like behavior at high currents.

Over a narrow range of parameters in the plastic regime, Bhattacharya
and Higgins observed jaggedness in the differential resistance.  This
corresponds to secondary peaks in addition to the main peak in the
differential resistance.  As the external magnetic field is varied,
the peaks can be made to appear and disappear, but for a given value
of magnetic field and for a given sample, the peaks are reproducible,
and thus act as ``fingerprints'' of the underlying pinning disorder.
Our model also exhibits such IV fingerprints.

\section{Some Comments on Conventional Theory}

The usual interpretation of the peak effect follows the original
proposition of Pippard\cite{pipard}, that softer systems are pinned
more strongly than more rigid ones.  This picture has been put on a
formal basis by Larkin and Ovchinnikov's theory of collective
pinning \cite{LO}. It was found that
critical current increases as the external magnetic field
increases near $H_{c2}$ because the FLL elastic moduli soften.
However, collective pinning theory doesn't explain the shape of the IV curve
outside of the elastic regime. Some advances have recently been
made by Le Doussal and Giamarchi to probe the transition to plastic
flow and other features \cite{doussal}.

MD simulations have indicated that, at
least for the two dimensional case, the pinning forces and IV
characteristics are determined by plastic deformations (tearing of the
FLL) which fall outside the region of validity of collective pinning
theory\cite{jensen1,jensen2}.  It is believed that the
 two dimensional elastic system is always 
unstable at onset, when a finite voltage first appears, and
flow must take place in the plastic regime.  These simulations can also
be interpreted as describing two dimensional cross-sections of the
three dimensional FLL in the plastic regime.  The scale of the MD 
simulations is that of the vortex cores, which is much smaller than
the London length, $\lambda$.

Shi and Berlinsky\cite{shi} presented a limited analytic treatment of
the way in which ``lattice defects alter the flow characteristics of
the lattice under the influence of an external drive''.  However both
the density of lattice defects  and the
IV relation were determined by numerical MD simulations.  These
simulations showed that at high currents where the IV relation is
linear, the defect density drops.  It increases sharply at lower
currents where the IV relation develops an S-shaped curve.

Further simulations have shown that the two dimensional plastic flow
takes place in terms of rivers of moving vortices separated by islands
where the vortices do not move\cite{noriscience,nori3}.  This
channel flow behavior has been observed experimentally using Lorentz
microscopy\cite{lorentz}.  Previous simulations of the coarse-grained
BP model revealed results consistent with these and also showed that
the channels can form a braided river which exhibit self-affine
(multifractal) behavior similar to fluvial braided rivers (see for
example Ref.~\onlinecite{braided-river-book}).

Nori and collaborators first studied, using MD simulations, flux
driven into superconductors with random pinning, with the driving
force solely due to the flux-density gradient.  They elucidated many
properties of the Bean state including the magnetic field profile,
magnetization hysteresis loops, critical currents, vortex avalanches,
and vortex rivers \cite{gdMD0,gdMD1,gdMD2,gdMD3,gdMD4,gdMD5}.  None of
these studies using an open system with an overall density gradient
reported the IV characteristics, though.

Recently, Nori {\it et al} \cite{nori1} have simulated the IV
curve as the FLL softens by varying the vortex interaction parameter
(see also Ref.~\onlinecite{chan3}).  As in all previous MD studies of
IV behavior of the FLL \cite{jensen1,jensen2,brass,gronbechjensen,nori1,nori2,kolton}, 
the vortices are contained in a
{\it periodic system, where they can neither enter nor
leave.} The initial condition is an ordered vortex lattice.  Motion
takes place via MD updates, with a {\it uniform force} applied to all
vortices.

The most important difference, besides the scale of the model, with
our description of the IV experiments, is that, in all of these MD
studies of the current-voltage relation, no overall vortex density
gradient can develop owing to the periodic boundary conditions.
 This makes the vortices travel perpetually around
the system and forces the paths to circle, which is unphysical, and
does not occur in the real system.  The actual physical situation is
an open system with flux pushed in and out, rather than a periodic
one.  The physical system adjusts its overall magnetic flux profile in
response to applied forces.  This is not possible in a periodic,
closed system.  Nevertheless, the basic result that the critical
current increases as the vortex interactions weakens is obtained.
However, all the IV curves measured via MD simulations of periodic
systems, fall on top of each other, or overlap, at high currents, for
different values of the vortex interaction strength.  This is
inconsistent with the experiments of Bhattacharya and Higgins (see
Fig.~1), and reflects the fact that the artificially periodic system
is not able to adjust its profile in response to applied forces.  An
even more significant, but related, difference is that the peak in the
differential resistance grows monotonically and gets sharper as the
resistance peak shifts to higher currents, the exact opposite of what
happens in experiments.

In short, the systematically changing morphology of the IV curves and
differential resistance for a soft FLL has not been theoretically
described.

\section{The Model}

\subsection{Motivation}
We take a different approach to modeling flux flow in
superconductors.  In order to describe collective transport in
disordered media, it is reasonable to consider coarse-grained models
where the details of the precise interactions between 
flux lines (or vortices) are lost
but the general effects of repulsive interaction
between granular or discrete objects, pinning, and over-damped
motion leading to stick-slip dynamics (tearing) are preserved.

The BP model is an interacting sand-pile model of vortices in a type-II
superconductor, where the ``sand'' grains, representing magnetic
vortices, repel each other.  It was originally motivated by the
observation of (possibly) self-organized critical\cite{soc} avalanches in field
ramping experiments\cite{field}, where the distribution of flux packages falling
into the interior coil of a hollow cylindrical superconductor were measured.
Actually, the similarity between the Bean state\cite{bean} (or vortex pile) and
sandpiles was first pointed out by de Gennes\cite{degennes}.  Later,
Vinokur, Feigel'man, and Geshkenbein\cite{VFG} suggested that
thermally induced flux creep would lead to a self-organized critical
state in a type II superconductor, as did Tang\cite{Tang}.  The key
observation is that flux line flow always takes place on a flux pile
which has an overall density gradient.  This pile may be in a
self-organized critical, or some other nontrivial state.

In addition to describing vortex avalanches in field
ramping experiments\cite{field}, the BP vortex model has also
been used to describe flux noise\cite{stroud}, vortex avalanches in the 
presence of a periodic, dense array of pinning centers\cite{cruz}, 
thermally activated flux creep\cite{mulet,newjensen1}, and magnetization 
loops\cite{newjensen2}. 

\subsubsection{Coarse-graining}

 Consider a
transverse two-dimensional slice of a superconducting slab at $T=0$.
The BP model\cite{kevin} results from a coarse-grained description to
the scale of the London length, $\lambda$, of the microscopic
dynamics, and incorporates the features that are essential to produce
the observed complex behavior: Repulsive interactions between
vortices, variations in the pinning potential, and variations in the
vortex density - all at the scale of $\lambda$.  One can imagine
imposing a grid of cells on the system; vortices in the model
correspond to a vortex number in an extended region of the actual
physical system.  Pinning in the model corresponds to a number of
point pins in an extended cell.  Each lattice site in the model can
hold many vortices, and can have a different, albeit quenched, pinning
potential, due to the underlying randomness in the positions and
strengths of the microscopic pinning centers.  The model allows many
vortices to interact with each other while maintain locality (at the
scale of $\lambda$) in interactions.  It is the only model of this
sort that has been proposed to describe flux dynamics in
superconductors. It enables
numerically studies, using ordinary
workstations, of the steady state and transient
properties of systems larger than
$(500\lambda)^2$ containing tens of millions of vortices. 

\subsection{Definition}

The BP model is defined as follows (see Fig.~3).  Consider a
two-dimensional honeycomb lattice \cite{lattice}
 where each cell, $x$, has three
nearest neighbors, and is occupied by an integer number of vortices,
$m(x)$.  The total energy of the vortex system includes the
repulsive pairwise interaction between the vortices 
and  the attractive interaction of vortices
with the pinning potential, ${\hat V}$.
  For a given configuration of vortex number
$\{m(x)\}$, the total energy of the system is:
\begin{equation}
H\bigl(\{m(x)\}\bigr)= \sum_{i,j}J_{ij}m(i)m(j) - \sum_i
{\hat V}_{\mbox{pin}}(i)m(i) \, \,  .
\end{equation}
Since the model describes a system
coarse-grained to the scale of the London
length, the repulsive
interactions $J_{ij}$ are short-ranged.  This usually includes
an on-site interaction, and a weaker nearest-neighbor interaction.

As in the microscopic case described by MD simulations,
the change in the total energy
of the model when moving a unit vortex from one site to a nearest
neighbor site is determined.   This yields
the force to move a unit vortex from $x$ to $y$, which is
\begin{eqnarray}
F_{x \rightarrow y} 
& = & V_{\mbox{pin}}(y) - V_{\mbox{pin}}(x) + [m(x) - m(y) -1] \nonumber \\
& & + r[m(x1) + m(x2) - m(y1) - m(y2)] \quad .
\end{eqnarray}
As indicated in Fig.~3, the nearest neighbor cells of $x$ are $y$,
$x1$, and $x2$, and the nearest neighbors cells of $y$ are $x$, $y1$,
and $y2$, and $0 \leq r<1$.  A slightly different implementation of
the disorder is used than before.  The normalized
pinning potential $V_{\mbox{pin}}(x)$
is a random number taken from a
uniform distribution in the interval between zero
and $V_{max}$.  In each time step, all cells of the lattice are
updated in parallel. A single vortex moves from a cell to a
neighboring cell if the force in that direction is positive, or equivalently
if the total energy of the system is lowered.  In Eq.~2, the units of force
on a vortex have been normalized so that the on-site term is unity.
Thus there are two dimensionless 
parameters remaining, $V_{max}$ and $r$.

\begin{figure}
\narrowtext
\epsfxsize=2.8truein
\centerline{ \epsffile{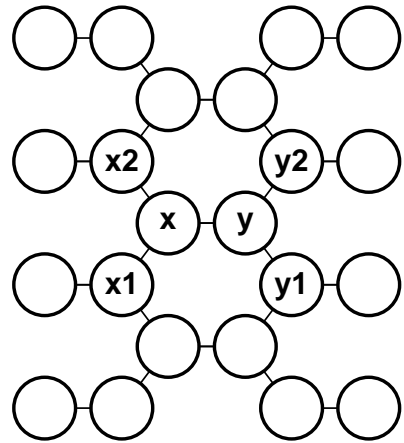} }
\caption{ 
The two-dimensional honeycomb 
lattice. Each cell, $x$, has three
nearest neighbors, and is occupied by an integer number of vortices, $m(x)$.
The force pushing a vortex from cell $x$ to
cell $y$ is calculated by taking the discrete gradient of the sum of
two potentials, one representing the repulsive interaction between vortices
occupying the same and nearest neighbor cells,
and the other representing the attractive
interaction between vortices and pinning centers.}
\end{figure}

Many alternatives exist to handle the situation when more than one
unstable direction appears for a vortex to move.  In the previous
implementation of the model, one unstable direction was chosen at
random.  In order to simplify the model and eliminate all potentially
spurious
sources of noise,  the most
unstable direction that has the largest force is chosen and the vortex
moves to that site.  This represents an extremal process.
In fact the entire model is now completely
deterministic, corresponding to a $T=0$ limit of the dynamics.

\subsection{The External Magnetic Field: Building a Vortex Pile}

Flux lines enter the superconductor from the edges, pushed in
by the external magnetic field.  This is represented by putting all sites on
the left edge of the model in contact with a  reservoir of
vortices at some potential, corresponding to the external magnetic field
on the left side of the sample.  The same is done for the right side of
the sample.  For simplicity, periodic boundary conditions are used for
the top and bottom.  If the two reservoirs are set equal, representing
embedding the sample in an external magnetic field, 
vortices enter the system generating the classic V-shaped flux density curve
as the external magnetic field is increased
(see below).

\subsubsection{Details about Boundary Algorithm}

In order to calculate the force on a vortex to move to or from a boundary
site, a special algorithm must be used because one of the nearest neighbor
sites of each boundary site is not on the lattice. The rule used here
simply assumes that the ``virtual''
 off lattice site neighboring each boundary site 
is occupied with an equal number of vortices as that boundary site.
More specifically, at the beginning 
of each lattice update, all of the sites on the boundary are set to
be occupied with
the same number of vortices. The lattice update then proceeds, during
which vortices can move off the boundary sites into the system, or from the 
system onto the boundary sites, thereby changing the number of 
vortices occupying
a boundary site. However, at the beginning of the next lattice update all
of the sites on the boundary are reset to their original value. Through
this process, vortices can be removed or added to the system. In general,
the left and right boundaries are held at different values. There is no
pinning at the boundary sites.  

\subsubsection{Details about Parallel Update}

An artifact of parallel updating is the existence of local
instabilities in which two, or more, vortices oscillate back and forth
between neighboring sites.  These local instabilities disappear if the
model is coarse-grained, because then the neighboring sites are
incorporated into a single one, and therefore are not important to the large
scale behavior of the system. The instabilities can be eliminated by
keeping track of the direction from which the last vortex moved onto
each site and always forbidding a return movement backwards in that
direction.  Otherwise, backward jumps are rare and therefore
disallowing them does not change the large scale behavior of the
system.  A similar rule applies to the boundary sites.
The advantage of the parallel update is that it is numerically more efficient,
than other update schemes
and does not introduce any uncontrollable spurious effects.

\subsection{The Transport Current: Shifting the Boundary Conditions}
\label{IV}

We consider an infinite slab of finite thickness in a parallel
applied magnetic field, carrying a current perpendicular to
the field.   Depending on the direction of the applied
current, the magnetic field on one side  of the superconductor
(e.g. the right hand side) will be decreased and the magnetic field on
the other side will be increased (e.g. the left hand side).
Assuming the applied magnetic field is sufficiently strong,   this
corresponds just to changing the heights of the magnetic flux pile on the
left and right edge, or a general kind of ``tilt'' of the pile.
See, for example, Ref. \cite{schmidt}.  Of
course, the internal currents and forces inside the superconductor
will readjust to accommodate this new boundary condition.
The flux lines  are considered to be perfectly stiff,
and described by a two-dimensional slice of the three dimensional
slab. 

\begin{figure}
\narrowtext
\epsfxsize=3.0truein
\centerline{ \epsffile{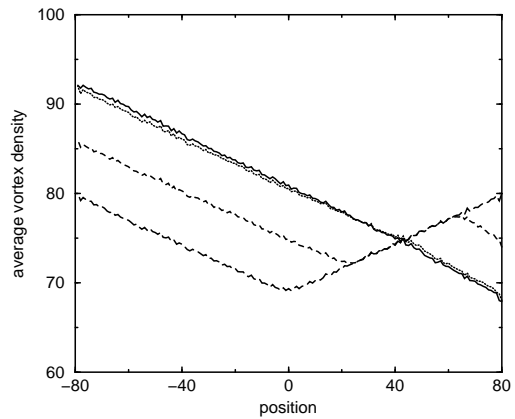} }
\caption{ 
Magnetic field profile as a function of distance  from the center of
a superconducting slab in the direction perpendicular to both
the external magnetic field and applied  current. 
The long dashed line is the stable profile when an external
field is applied, but no applied current, $I$. The short dashed, and
dotted lines are the stable hysteretic profiles resulting from increasing 
applied current. The solid line is the profile just above the onset
of vortex motion.}
\end{figure}

An applied magnetic field in the geometry
described above will produce the well known ``V
profile'' of the magnetic flux density discussed, for example, in
Orlando\cite{orlando} and observed in many experiments, such
as those of Behnia, {\it
et al}\cite{behnia}.  This is shown in Fig.~4.
The actual magnetic profile depends on the
history of the sample and how magnetic field has been applied in the
past.  Applying a finite current shifts the boundary conditions resulting in
a hysteretic profile, also shown in Fig.~4.

\subsubsection{Onset}

Eventually, as the applied current is increased further, a critical
current is reached, where the shift of the boundary conditions is so
large that steady vortex flow occurs down the gradient
spannning the entire sample.  A profile
just above the critical current is also shown in Fig.~4.  Note that
this profile also retains some hysteretic properties.  For example, it
has a bump corresponding to where the increasing and decreasing
portions of the magnetic field profile merged as the external current
was increased above the threshold.  This bump disappears if 
the boundary conditions on
the sample are shifted further and then lowered back
to  the previous value.

\subsection{Making IV Measurements}

The IV characteristic is determined by the relation between applied
current and the vortex flow, which induces a voltage.  To our
knowledge, this type of numerical measurement, made by shifting the
boundary conditions on the Bean state, has not been investigated
before.  Here the IV characteristic is simply the relation between
the magnitude of the shift (representing an applied current) and the
average flow rate of vortices (representing a voltage) when the
critical current (tilt) is exceeded.  In order not to confuse the
reader we use the term current to refer to the applied electrical
transport 
current and the term flow to refer to the motion of magnetic vortices.

In general, the boundary sites in the model can be set to any real
value, including non-integer values. This is because vortex number on
the boundary sites describes the external magnetic field density
at the boundary of the sample. However, only integer units of
magnetic flux can enter
the interior of the system, and thus only integer numbers of vortices
occupy interior lattice sites. Obviously, the magnitude of the
difference between the heights of the left and right boundaries, can
also be set to non-integer values.
However, if the boundary heights are shifted by less than whole integers,
steps can appear in the IV data. These steps are caused
by the fact that only whole numbers of vortices can occupy interior lattice
sites. Since one vortex unit in the model represents many actual physical
vortices, this is to some extent an
unphysical artifact of the model, and
the discreteness effect should be less apparent in
experiments. To eliminate this effect, all of the IV data presented in this 
paper
were calculated by shifting the boundary heights by only integer numbers of
vortices.

In simulations of the model, there is almost no backward movement of vortices.
Thus, the vortex flow can be determined by measuring the
average number of vortices moving per lattice update (the average
activity). Furthermore, the velocity of
each moving vortex is one lattice site per update. Thus,
the experimentally measured voltage, which is equal to the
amount of moving flux times the velocity of that flux, is
therefore proportional to  
the average vortex activity.

\begin{figure}
\narrowtext
\epsfxsize=3.0truein
\centerline{ \epsffile{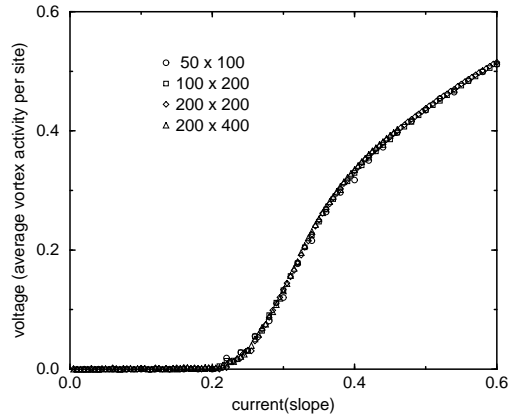} }
\caption{ 
Finite-size scaling plot of IV measurements from simulations of the cellular 
model. 
System sizes are shown in the legend.
}
\end{figure}

IV data from simulations of different size systems collapse nicely in a scaling
plot, Fig.~5, if the voltage is measured as vortices moving per lattice update
per lattice site (the average activity per site), 
and the current is measured as the average slope
across the system (the magnitude of the height difference between
the left and right boundaries divided by the length of the system). 
In the following sections, IV data is presented in these
scaling units.
In Fig.~5, the IV data was produced by repeatedly increasing the height of the
left boundary by 1 vortex and lowering the height right boundary by
1 vortex. The vortex interaction strength was $r=0.1$. 

The IV data presented in the following sections was calculated in a
similar fashion. All of the data is for systems of size 200x400.  Each
IV data point was calculated by first shifting the left boundary
height up 1 vortex and shifting the right boundary height down 1
vortex.  As for Fig.~5, the lattice was updated 20000 times to
eliminate transient behavior, and finally the lattice was updated
another 20000 times during which the average vortex activity was
measured.

\section{Collective Transport Behavior of the Model}

The IV relation was measured for different values of the parameters
 and for different system sizes.  Our main result is
shown in Fig.~6, where the parameter $V_{max}=5$  and  the parameter
$r$ is varied.  The parameter $r$  represents the strength
of repulsion between vortices at nearest neighbor cells (of size $\lambda$).

\begin{figure}
\narrowtext
\epsfxsize=3.0truein
\centerline{ \epsffile{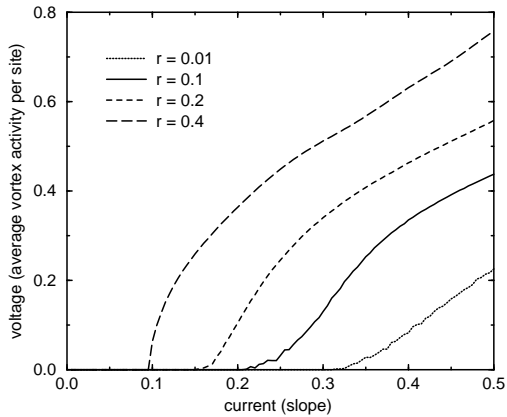} }
\caption{ IV measurements from numerical simulations, for four different value
of the vortex interaction strength, $r$.}
\end{figure}

The first result is that as $r$ decreases, the critical current, which
is the slope of the pile where vortices first start to flow,
increases.  Clearly, applying an increased $r$ to the pile in the
steady state lowers the slope of the pile since formerly stable local
slopes will now become unstable due to the increased repulsion between
vortices at neighboring sites.  Thus, fixing all other parameters,
 we can identify the parameter $r$
in the model as a way of controlling the critical current, or slope of
the pile.  In the superconductor, the critical current can be
controlled by the applied magnetic field.  In the peak effect regime,
it turns out that
increasing the applied magnetic field leads to a softer FLL
 and thus a higher critical current.  Therefore, the regime
where the critical current is an increasing function of the applied
magnetic field is represented in our model by a critical current which
is a decreasing function of the parameter $r$.  This is made evident
by comparing Figs.~1,2 and Figs.~6,7.

The second result is that, except for the largest $r$, all of the IV
curves have a characteristic S shape.  They start out at some $I_c$
increasing concave upward and then bend over saturating to a finite
slope at large currents.  The IV relations for a given realization of
disorder, however, do not overlap at high currents, unlike the results
obtained with previous MD simulations \cite{nori1}.  Again, all of
this agrees with the experimental results.

\begin{figure}
\narrowtext
\epsfxsize=3.0truein
\centerline{ \epsffile{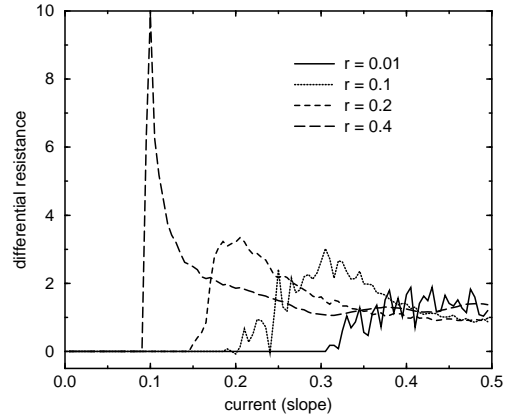} }
\caption{ 
Differential resistance measurements from numerical simulations.
These results are calculated by numerical differentiation of the IV results
shown in Fig.~6.}
\end{figure}

In Fig.~7, the differential resistance $dV/dI$ is calculated by taking
numerical derivatives of the curves shown in Fig.~6.  At the largest $r$ 
value, the resistance peak
is very large.  As the parameter
$r$ decreases, the peak moves to higher currents
and decreases in magnitude.  In fact the same behavior is observed in
experiments as the external magnetic field approaches $H_{c2}$.  As
mentioned before, the MD simulations give the opposite result of a
peak that increases in magnitude as it shifts to higher currents\cite{nori1}.

In the actual experiments, at fields below the peak effect regime, the vortex
flow is believed to be elastic and there is no observed peak in the
differential resistance.  As the magnetic field increases a peak in
the differential resistance starts to develop, which then reaches a
maximum, decreasing again for larger fields.  Our simulations appear
to describe the experiments once the peak in differential resistance
has reached its maximum.  

In order to obtain an elastic regime, we could
consider a three dimensional coarse-grained model of repelling
flux lines rather than point vortices representing a two-dimensional
cross section of that system.  Work is in progress along those lines.

The model discussed here does
not describe the behavior of the superconductor in fields greater than
that which gives the largest critical current, where the critical
current decreases as the magnetic field increases.  This may be
due to the fact that we take the depairing current to be strictly
infinite (i.e. $j_c/j_0=0$)
and there is no transition to a non-superconducting state in
the model presented here.

Note that the differential resistance curves for small $r$ contain
secondary peaks, in addition to the main peak.  This is similar to the
jaggedness or ``fingerprint'' found in experiments.  This jaggedness
in our results occurs in the filamentary channel 
regime, discussed below, and
is due to filaments opening and closing as the applied current increases.

\subsection{Relation to River Morphology}

So far we have only characterized the model using measurements
analogous to those that experimentalists typically have available.
However, numerical simulations can also provide a great deal of easily
accessible information about the morphologies of flow patterns
associated with different configurations.  In fact, different
morphologies of vortex flow patterns have been observed using MD
simulations \cite{gdMD3,gdMD4,gdMD5,jensen1,gronbechjensen,chan3,moon}.  To
understand the nature of the differences in the shape of the IV
curves, we have examined how the flow morphologies change as one
increases the applied current for a system that has an S-shaped IV
curve, and also for the singular IV curve that occurs at large $r$.

The paths that the vortices take as they cross the sample can be
determined by measuring the average activity at each site.  For
example, Fig.~8 shows a series of grey scale images of the vortex flow
patterns. These images represent ``time-lapsed photographs'' of the
vortex activity. The different images in Fig.~8 show the vortex flow
patterns for the case $r=0.1$ as a function of increasing external
current, $I$. In these images, sites with no vortex activity are
blank, sites with an activity greater than or equal to 0.5 (one vortex
moves every other lattice update) are black, and sites with activity
between 0 and 0.5 are indicated by grey dots with a darkness
proportional to their activity.  These patterns are fixed
and do not change in time with fixed external driving conditions.

The nature of the flow can be quantified by the distribution of
activity at the different lattice sites.  Figure 9 shows histograms of
the activity corresponding to the images of Fig.~8.  These histograms
are constructed with 1000 bins and normalized so that the area under
the curve is equal to one. Note that there are peaks corresponding to
zero activity not included in the figure.  Those peaks at zero
activity decrease in size as the current is increased.

As can be seen in Figs.~6 and 7, the IV curve for $r=0.1$ is
S-shaped. Figure 8a shows the vortex flow pattern just above threshold
for that case. The vortex flow takes place only on a single
filamentary string with some small side branches. This behavior is
also evident in the histogram Fig.~9a, which shows a small number of
isolated peaks.  As the tilt of the pile increases the number of
filaments of the vortex flow increases, and they begin to merge. This
process can be seen in Figs.~8b and 8c. The corresponding histograms
of activity shown in Figs.~9b and 9c indicate an increasing number of
peaks, and the development of a continuous distribution. 
{\it During this
merging process, while the vortex flow remains filamentary, the IV
curve remains concave upward, and the differential resistance rises.}

Eventually the filaments of vortex flow 
merge to form a braided river, as shown in Fig.~8d.  This occurs
around the peak in differential resistance, corresponding to a change
in curvature of the IV relation. In this case,
the corresponding histogram of activity 
(Fig.~9d) shows a continuous distribution of activity peaked at zero.
{ \it Thus the peak in differential resistance signals a change in the
underlying vortex flow morphology from filamentary strings to a
braided river.} This appears to be consistent with the observation
of Kolton {\it et al} that at the peak in differential resistance
``all the vortices are moving in a seemingly isotropic channel
network with maximum interconnectivity'' \cite{kolton}.

As the tilt of the pile is increased even further, the vortex flow
becomes spatially more and more uniform, as the braided river floods.
This uniform flow region corresponds to linear, or Ohmic
behavior. These results can be seen in the river flow pictures in
Figs.~8e and 8f.  The corresponding histograms shows a peaked
continuous distribution, which has separated from zero activity.  As
the tilt increases further the continuous distribution of activity
obtains an increasing mean and narrowing width.
{\it The transition to Ohmic behavior comes about as the braided river
floods at higher flow rates than can be supported by such a structure.}

\twocolumn[\hsize\textwidth\columnwidth\hsize\csname
@twocolumnfalse\endcsname 
\begin{figure}
\epsfxsize=6.0truein
\centerline{ \epsffile{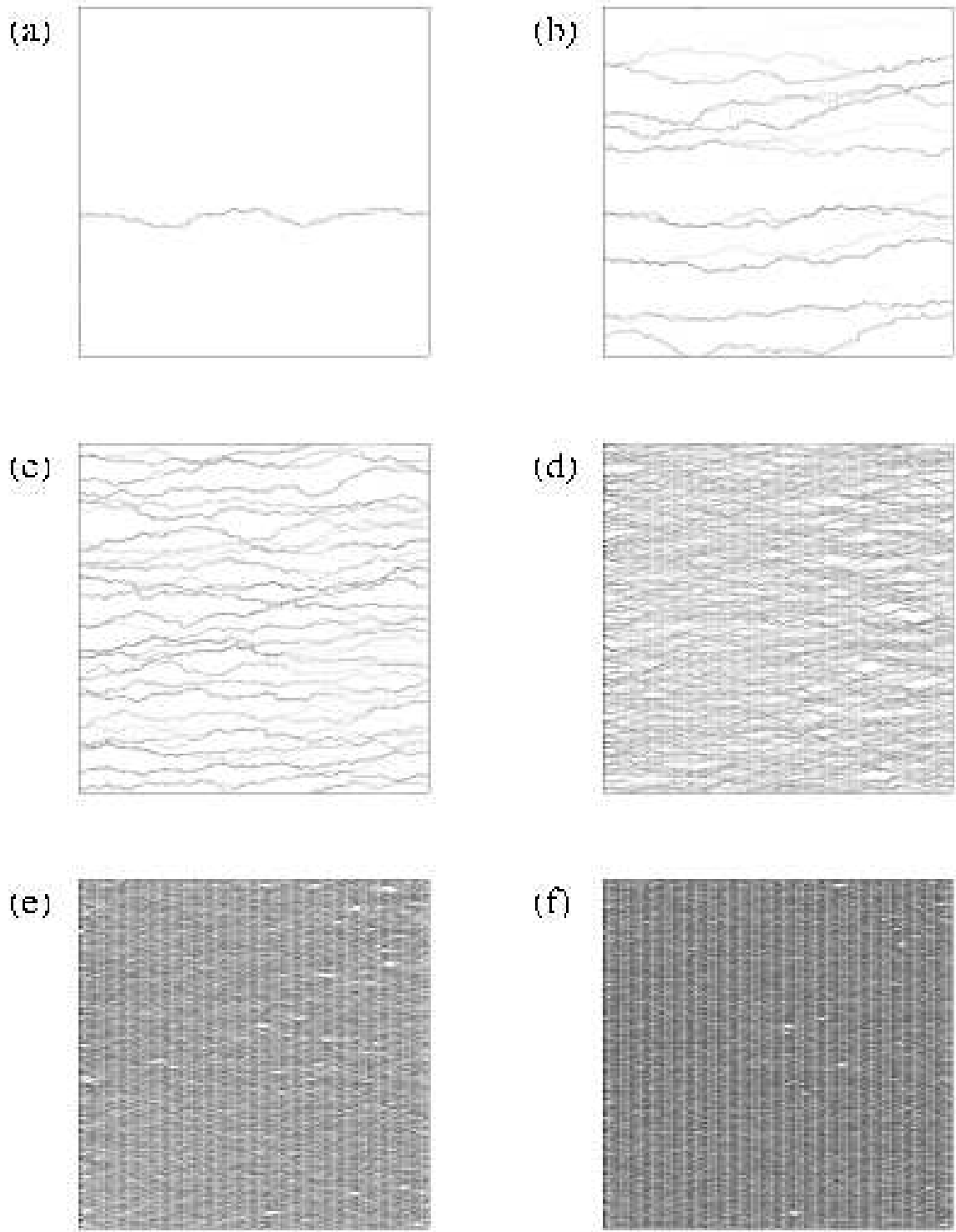} }
\caption{ 
Vortex flow patterns at different values of the external
transport current for a vortex interaction strength of
$r=0.1$ with a corresponding S-shaped IV curve. The current in
each case (measured as the slope of the system) is: (a) 0.21,
(b) 0.225, (c) 0.25, (d) 0.30, (e) 0.375, and (f) 0.45.}
\end{figure}
\vskip2pc]

\twocolumn[\hsize\textwidth\columnwidth\hsize\csname
@twocolumnfalse\endcsname 
\begin{figure}
\epsfxsize=6.0truein
\centerline{ \epsffile{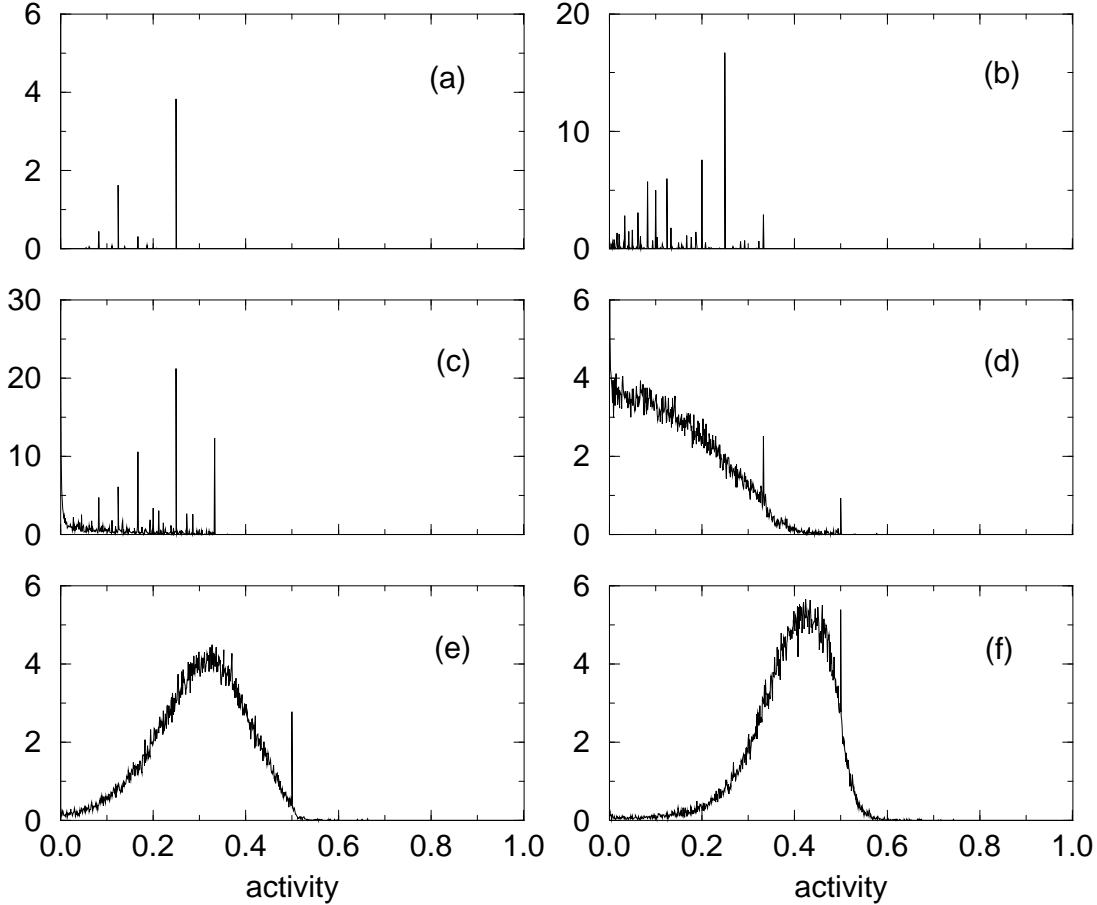} }
\caption{ 
Histograms of the site activity corresponding to the vortex flow
patterns shown in Fig.~8. Not visible in the figures are the
peaks at zero activity. The size of those
peaks are: (a) 993, (b) 920, (c) 735, (d) 95, (e) 14, and (f) 7.6 .
}
\end{figure}
\vskip2pc]

A similar progression of flow morphologies, from isolated filamentary
channels, to a braided river, to a flooded river with uniform flow,
occurs for other values of $r$ with S-shaped IV curves. However, a
different scenerio occurs at large values of $r$ where the IV curve is
not S-shaped, but is instead always concave downward from the onset of
vortex motion. The IV curve for $r=0.4$ in Fig.~6 is an example.  We
will refer to this as the critical curve; its functional form is
discussed in the next subsection. The differential resistance for the
critical curve is discontinuous, as can be seen in Fig.~7.

For the critical curve, there is no filamentary region of
vortex flow. {\it Instead, right at the onset of vortex motion, the flow
 has the structure of a braided river.} This can be seen in
Fig.~10a, and in the corresponding activity histogram in Fig.~11a
which shows a continuous distribution of activity.
(Figures 10 and 11 were produced in the same manner as Figs.~8 and 9,
respectively.)
As the tilt of the pile is increased even further,
the flow again increases and becomes more spatially uniform,
similar to behavior of the S-shaped IV case after the peak in
the differential resistance, where the braided river floods.
 This behavior can be seen in the river 
flow pictures in Figs.~10b, 10c and 10d, and the corresponding histograms 
in Figs.~11b, 11c and 11d.  Since the critical curve has a diverging
differential resistance at onset,
this further supports our observation that
the peak in the differential resistance is associated with a braided
river structure.

\twocolumn[\hsize\textwidth\columnwidth\hsize\csname
@twocolumnfalse\endcsname 
\begin{figure}
\epsfxsize=6.0truein
\centerline{ \epsffile{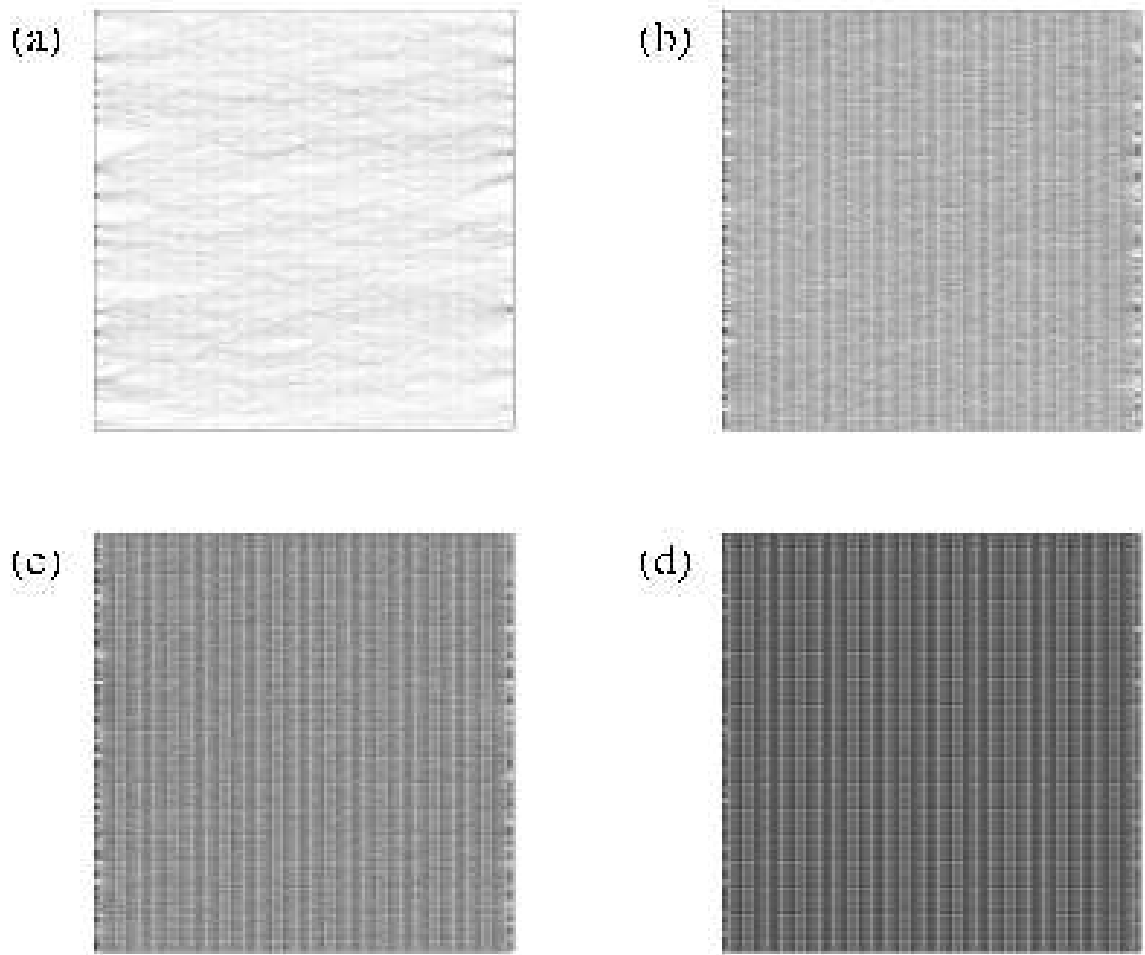} }
\caption{ 
Vortex flow patterns at different values of the external
transport current for a vortex interaction strength of
$r=0.4$ with a corresponding critical IV curve. The current in
each case (measured as the slope of the system) is: (a) 0.10,
(b) 0.15, (c) 0.20, and (d) 0.40.}
\end{figure}
\vskip2pc]

\twocolumn[\hsize\textwidth\columnwidth\hsize\csname
@twocolumnfalse\endcsname 
\begin{figure}
\epsfxsize=6.0truein
\centerline{ \epsffile{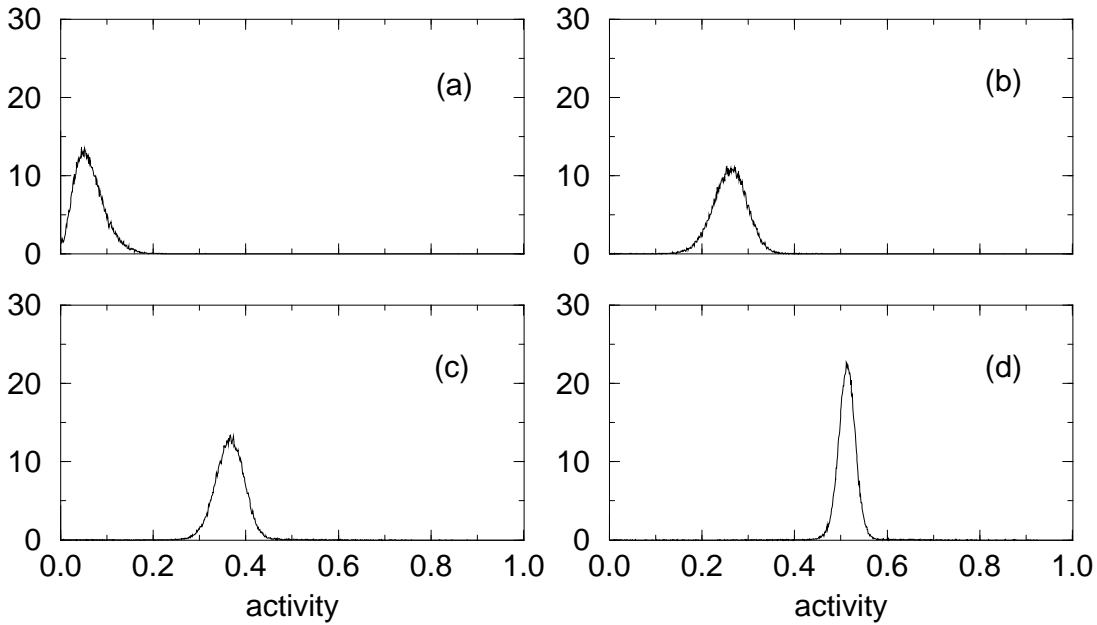} }
\caption{ 
Histograms of the site activity corresponding to the vortex flow
patterns shown in Fig.~10. Not visible in the figures are the
peaks at zero activity. The size of those
peaks are: (a) 15.7, (b) 5.6, (c) 4.4, and (d) 3.3 . }
\end{figure}
\vskip2pc]

\subsection{Scaling of the Critical IV Curve}

The critical  IV curve for $r=0.4$ is well described by 
$$
V \sim (I-I_c)^\beta \quad ,
$$
with $\beta = 0.6 \pm 0.1$, as shown in Fig.~12. For that
measurement, the critical current, $I_c,$ was measured to be
$0.095$, which was the largest current found with zero voltage.
Since the current was sampled at values spaced by $0.005$,
the uncertainty in $I_c$ is approximately $\pm 0.005$.
Varying the value of $I_c$ over that range changes the value of $\beta$ that
best fits the data near the onset of vortex motion, and allows 
an estimate of the error on $\beta$.

The exponent $\beta$ can be related via scaling arguments to the
exponents characterizing the distribution of avalanches in the
self-organized critical state.  In the self-organized critical state,
the average flow rate, $V$, of vortices is controlled rather than the
overall slope of the system.  The scaling argument is similar to that
used in Corral and Paczuski \cite{alvaro} to describe the transition
from avalanche to continuous flow in the one dimensional Oslo model
\cite{alvaro}.  The excess slope above the critical slope at onset is
$\Delta m = I-I_c= (\Delta N)L^3$, where $\Delta N$ is the excess
number of vortices in the system, and $L$ is the system size.  If
vortices are added very slowly then there will be distinct avalanches
separated by intervals of no activity.  In that regime, superposition
applies and $\Delta N \sim V L^z$, where $L^z$ is the cutoff in the
duration of the avalanches.  In the rapidly driven regime, the
avalanches overlap, and the excess slope becomes independent of system
size, thus $\Delta m \sim V^{1/\beta}$.  These two limits can be
combined into a single cross-over scaling function:
\begin{equation}
\Delta N \sim L^a f(VL^x) \quad .
\end{equation}
The exponent $x$ in this expression measures the average duration of
avalanches $\langle t\rangle \sim L^x$.  Since avalanches in this case
come about from adding an entire row of $L$ vortices to the system,
the average size of avalanches, measured in terms of the number of
topplings is $\langle s\rangle \sim L^2$, rather than $\langle
s\rangle \sim L$, as in the Oslo model.  Obviously, on average
each site in the system topples once when a row of vortices is added
in the self-organized critical state.  Using this result together
with conservation of probability gives a scaling
relation between $x$ and $z$;
$$
x=z+2-D \quad .
$$
Combining all these results gives
$$
\beta = {x \over x+2-z }= {z+2-D \over 4-D} \quad .
$$
Using the exponent values $z=1.5$ and $D=2.7$ obtained in
Ref. \cite{kevin} gives $\beta=0.6$, in good agreement with the numerical
result presented in Fig.~12.

\begin{figure}
\narrowtext
\epsfxsize=3.0truein
\centerline{ \epsffile{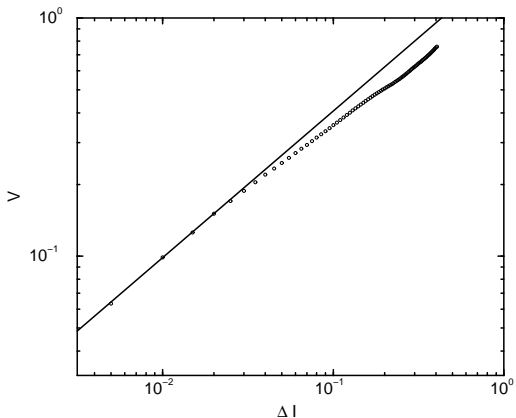} }
\caption{ 
Double logarithmic plot of voltage verses current minus critical current
for $r=0.4$.The straight line shown
has a slope of $0.615$. }
\end{figure}

\section{Summary}
We have studied the nonlinear current-voltage characteristics of flux
flow in type II supeconductors with random pinning
using a simple, cellular model.  As in
the physical system, vortices flow down a flux density
gradient. Because the coarse-grained model does not include any
degrees of freedom at scales much smaller than the London length,
$\lambda$, it can not describe structural ordering or disordering of
the flux line lattice. Despite this fact, simulations of the model
reproduce many of the empirically observed transport characteristics
of plastic flux flow in type II superconductors.

In particular, our results reproduce many of the features attributable to
plastic flow of the FLL in the ``peak regime''.  By weakening the
vortex interaction strength, we find an increase of the critical
current and a falling of the magnitude of the peak in the differential
resistance.  The model also exhibits IV fingerprints, and crossover to
Ohmic or linear behavior at high currents. Also, the IV curves for
different vortex interaction strengths do not merge at large external
currents. All of these features are completely consistent with
experimental results.  However, presumably due to the
two-dimensionality of the model, the elastic behavior of the FLL is
not reproduced.

The success of these efforts to describe the plastic transport
behavior of magnetic flux in superconductors
with a coarse-grained cellular
model suggests a possibly generic explanation for and ubiquity of
plastic flow phenomena observed in superconductors.  It may not depend
on the degree of disorder or defects in an underlying microscopic flux
line lattice. Instead the behavior may be common to driven repulsive
particle systems in a disordered media.  The varieties of plastic flow
behaviors in the model studied here result from the changing
morphologies of the vortex flow pattern down a vortex density
gradient, self-organizing into distinct large scale patterns.  These
include isolated filaments, which merge at higher flow rates to give a
braided river, and lead to uniform flow,
or Ohmic behavior,
at the highest flow rates.
The filamentary structure is
associated with a concave IV characteristic; the braided river
structure is associated with the peak in the differential resistance;
the change to Ohmic behavior comes about as the braided river floods
as it can not support a higher level of flow.

\acknowledgements
This work was supported by the NSF, grant
\#DMR-0074613, and also by the Texas Center for Superconductivity.
E. Altshuler thanks the World Laboratory for Pan-American Collaboration
in Science and Technology for financial support.
We thank Per Bak for comments on the manuscript, and C.S. Ting and
G.F. Reiter
for informative conversations on superconductivity.


\begin{references}

\bibitem{vort-rev} A. M. Campbell and J. E. Evetts, {\it
Critical Currents in Superconductors,} (Taylor \& Francis, London,
1972). 

\bibitem{vort-rev2} G. Blatter, M. V. Feigel'man, V. B. Geshkenbein,
A. I. Larkin and V. M. Vinokur, Rev. Mod. Phys. {\bf 66}, 1125
(1994).

\bibitem{HB96}
M. J. Higgins and S. Bhattacharya, Physica C {\bf 257}, 232 (1996). 

\bibitem{kevin}
K. E. Bassler and M. Paczuski, Phys. Rev. Lett. {\bf 81}, 3761 (1998).


\bibitem{gdMD0}
R. Richardson, O. Pla, and F. Nori, Phys. Rev. Lett. {\bf 72}, 1268 (1994).

\bibitem{gdMD1}C. Reichhardt, C. J. Olson, J. Groth, S. Field, and F. Nori, 
Phys. Rev. B {\bf 52},  10441 (1995).

\bibitem{gdMD2}
C. Reichhardt, C. J. Olson, J. Groth, S. Field, and F. Nori, Phys. Rev. B 
{\bf 53}, R8898 (1996).

\bibitem{gdMD3}
C. Reichhardt, C. J. Olson, and F. Nori, Phys. Rev. B {\bf 56}, 6175 (1997).

\bibitem{gdMD4}
C. J. Olson, C. Reichhardt, and F. Nori, Phys. Rev. Lett. {\bf 80}, 2197 (1998).

\bibitem{gdMD5}
A. Mehta, C. Reichhardt, C.J. Olson, and F. Nori, Phys. Rev. Lett. {\bf 82}, 3641 (1999). 

\bibitem{shi}
A. C. Shi and A. J. Berlinksy, Phys. Rev. Lett. {\bf 67 }, 1926 (1991).

\bibitem{jensen1}
 H.J. Jensen, A. Brass, and A.J. Berlinsky,  Phys. Rev.
Lett. {\bf 60}, 1676 (1988)

\bibitem{jensen2}
 H.J. Jensen, A. Brass, Y. Brechet, and A.J. Berlinsky,
Phys. Rev. B {\bf 38}, 9235 (1988).

\bibitem{brass}A. Brass. H.J. Jensen, and A.J. Berlinsky,
Phys. Rev. B {\bf 39}, 102 (1989).

\bibitem{gronbechjensen}
N. Gronbech-Jensen, A.R. Bishop, and D. Dominguez,
Phys. Rev. Lett. {\bf 76}, 2985 (1996).

\bibitem{nori1}
C. J. Olson, C. Reichhardt, and F. Nori,
Phys. Rev. Lett. {\bf 81}, 3757 (1998).

\bibitem{nori2}
C. Reichhardt, C. J. Olson, and F. Nori,
Phys. Rev. B {\bf 58}, 6534 (1998).

\bibitem{kolton} 
A.B. Kolton, D. Dominguez,  and N. Gronbech-Jensen,
Phys. Rev. Lett. {\bf 83}, 3061 (1999).

\bibitem{braidedrivers} K. E. Bassler, M. Paczuski, and G. F. Reiter,
Phys. Rev. Lett. {\bf 83}, 3956 (1999).

\bibitem{field} S. Field, J. Witt, F. Nori and X. S. Ling,
Phys. Rev. Lett. {\bf 74}, 1206 (1995);
C. Heiden and G. I. Rochlin,
Phys. Rev. Lett. {\bf 21}, 691 (1968).

\bibitem{lorentz} 
T. Matsuda, K. Harada, H. Kasai, O. Kamimura and A. Tonomura,
Science {\bf 271}, 1393 (1996).

\bibitem{bh-prl93} S. Bhattacharya and M.J. Higgins, 
Phys. Rev. Lett. {\bf 70}, 2617
(1993).

\bibitem{bhfingerprints}
S. Bhattacharya and M. J. Higgins, Phys. Rev. B {\bf 52}, 64 (1995). 

\bibitem{marley} A. C. Marley, H. J. Higgins and S. Bhattacharya,
Phys. Rev. Lett. {\bf 74}, 3029 (1995).

\bibitem{pipard} A. B. Pippard, Philos. Mag. {\bf 19}, 217 (1969).

\bibitem{LO} A.I. Larkin and Yu. N. Ovchinnikov, J. Low Temp. Phys. {\bf 34}, 
409 (1979).

\bibitem{noriscience} F. Nori, Science {\bf 276}, 1373 (1996).

\bibitem{nori3}
C. J. Olson, C. Reichhardt, and F. Nori,
Phys. Rev. Lett. {\bf 80}, 2197 (1998).
 
\bibitem{braided-river-book} {\it Braided Rivers,} J.L. Best (ed.),
(American Association of Petroleum Geologists, 1993).

\bibitem{chan3}M.C. Faleski, M.C. Marchetti, and A.A. Middleton,
Phys. Rev. B {\bf 54}, 12427 (1996).

\bibitem{doussal} P. Le Doussal and T. Giamarchi, Phys. Rev. B {\bf 57},
11356 (1998).

\bibitem{soc}
P. Bak, C. Tang, and K. Wiesenfeld, Phys. Rev. Lett. {\bf 59}, 381
(1987); Phys. Rev. A. {\bf 38}, 364 (1988); for a review see P. Bak,
 {\it How Nature Works: The Science of
Self-Organized Criticality,} (Copernicus, New York, 1996).

\bibitem{bean} C. P. Bean, Rev.\ Mod.\ Phys.\ {\bf 36}, 31 (1964).

\bibitem{degennes} P. G. de Gennes, {\it Superconductivity of
Metals
 and Alloys}, (Benjamin, New York, 1966).

\bibitem{VFG} V. M. Vinokur, M. V. Feigel'man, and V. B. Geshkenbein,
Phys. Rev. Lett. {\bf 67}, 915 (1991).

\bibitem{Tang} C. Tang, Physica A {\bf 154}, 315 (1993).

\bibitem{stroud}
G. Mohler and D. Stroud, Phys. Rev. B {\bf 60}, 9738 (1999).

\bibitem{cruz}
R. Cruz, R. Mulet, and E. Altshuler,
Physica A {\bf 275}, 15 (2000).

\bibitem{mulet}
R. Mulet, R. Cruz, and E. Altshuler, cond-mat/9912103.

\bibitem{newjensen1} H.K. Jensen, and M. Nicodemi, cond-mat/0006491.

\bibitem{newjensen2} M. Nicodemi, and H.K. Jensen, cond-mat/0007028.


\bibitem{lattice} The choice of the lattice is somewhat arbitrary.
We can also choose a square lattice, for example.

\bibitem{schmidt}V.V. Schmidt, {\it The Physics of Superconductors,}
(Springer, Berlin, 1997).

\bibitem{orlando} T.P. Orlando and K.A. Delin, {\it Foundations of Applied 
Superconductivity,}
(Addison-Wesley, New York, 1991).

\bibitem{behnia}
K. Behnia, C. Capan, D. Mailly, and B. Etienne,
J. Low Temp. Phys. {\bf 117}, 1435 (1999);
Phys. Rev. B {\bf 61}, R3815 (2000).

\bibitem{moon}K. Moon, R.T. Scalettar, G.T. Zimanyi, Phys. Rev. Lett.
{\bf 77}, 2778 (1996).

\bibitem{alvaro}
A. Corral and M. Paczuski, Phys. Rev. Lett.
 {\bf 83}, 572 (1999).

\end{references}
\end{document}